\documentclass[12pt]{article}
\usepackage{latexsym}
\usepackage{amssymb}
\usepackage{amsfonts}
\usepackage{pslatex}
\usepackage{graphicx}

\catcode`\@=11
\def\numberbysection{\@addtoreset{equation}{section}
        \def\theequation{\thesection.\arabic{equation}}}

\def\be{\begin{equation}}
\def\ee{\end{equation}}
\def\ba{\begin{eqnarray}}
\def\ea{\end{eqnarray}}

\def\ov{\overline}

\def\Z{\mathbb{Z}}
\def\C{\mathbb{C}}
\def\RR{\mathbb{R}}
\def\nl{\nonumber \\}
\def\winf{W_\infty}

\def\ra{\rangle}

\def\la{\langle}
\def\de{\partial}
\def\wt{\widetilde}
\def\wh{\widehat}
\def\Tr{{\rm Tr}}
\def\dag{\dagger}

\def\a{\alpha}
\def\b{\beta}
\def\g{\gamma}

\def\D{\Delta}
\def\d{\delta}
\def\e{\epsilon}
\def\eps{\varepsilon}

\def\th{\theta}

\def\l{\lambda}
\def\L{\Lambda}

\def\p{\pi}

\def\r{\rho}
\def\s{\sigma}

\def\f{\varphi}

\def\c{\chi}
\def\w{\omega}

\numberbysection

\begin{document}
\begin{titlepage}
\begin{center}

\hfill  \quad DFF 420-09-04 \\

\vspace{1cm} 

{\Large \bf Matrix Model Description } \\ 
\vspace{.3cm}
{\Large \bf of Laughlin Hall States }\\ 

\vspace{1cm}

Andrea CAPPELLI\footnote{
Present address: TH-PH Division, CERN, CH-1211 Geneve 23, Switzerland.}
,\ \ Mauro RICCARDI \\
\medskip
{\em I.N.F.N. and Dipartimento di Fisica}\\
{\em  Via G. Sansone 1, 50019 Sesto Fiorentino - Firenze, Italy} \\
\end{center}

\vspace{.5cm}
\begin{abstract}
We analyze Susskind's proposal of applying the non-commutative 
Chern-Simons theory to the quantum Hall effect.
We study the corresponding regularized matrix
Chern-Simons theory introduced by Polychronakos.
We use holomorphic quantization and perform a change of matrix variables 
that solves the Gauss law constraint. 
The remaining physical degrees of freedom are the complex eigenvalues
that can be interpreted as the coordinates of electrons 
in the lowest Landau level with Laughlin's wave function.
At the same time, a statistical interaction is generated 
among the electrons that is necessary to stabilize the ground state.
The stability conditions can be expressed as the highest-weight conditions
for the representations of the W-infinity algebra in the matrix
theory. This symmetry provides a coordinate-independent 
characterization of the incompressible quantum Hall states.
\end{abstract}

\vfill
\end{titlepage}
\pagenumbering{arabic}


\section{Introduction}
In 2001 Susskind wrote an interesting paper \cite{susskind} 
where he suggested that the non-commutative Chern-Simons theory could
describe the Laughlin incompressible fluids \cite{laugh}  in
the fractional quantum Hall effect \cite{prange}.
His work was inspired by the analogies between the physics of electrons
in a strong magnetic field
and the properties of D branes in string theory \cite{susskind2}. 
In his argument, Susskind derived the semiclassical theory of the
incompressible fluids in a magnetic field and showed that 
it corresponds to the non-commutative theory in the limit of 
small $\th$ (high density); 
then, he proposed the fully quantum, non-commutative theory for 
the Laughlin Hall states made of discrete electrons. 

Several authors \cite{poly1,poly3,heller,karabali,jackiw,hansson,barbon} 
have then addressed the question whether 
the non-commutative Chern-Simons theory really realizes the Laughlin wave 
functions and the physics of anyons \cite{wilczek}.  
If the answer is positive, this theory would provide a promising approach 
for understanding the open issues of the quantum Hall effect, in
the form of an ``effective non-relativistic theory''.
Moreover, it would offer an interesting physical realization of the new
mathematical structures of non-commutative field theory \cite{NCFT} and 
non-commutative geometry \cite{ncgeom}.

Effective field theory descriptions of the quantum Hall effect 
has been extensively developed in the past years using
ordinary Chern-Simons theories \cite{froelich}, that correspond 
to conformal field theories of the low-energy excitations at 
the edge of the sample \cite{wen}.
These approaches have been rather successful and have been
experimentally confirmed \cite{saleur}. However, they
present some limitations, such as the need of several conformal
theories to describe the whole set of observed Hall plateaus and the presence
of slightly different proposals \cite{wen}\cite{minmod} for describing the
less understood Jain plateaus \cite{jain}.
Moreover, these effective descriptions do not incorporate the microscopic
physics to understand the ``universality'' of the Laughlin wave function
\cite{fradkin}, the Jain ``composite fermion'' transformation\footnote{
See the Refs. \cite{cftheories} for the theories of composite fermions.} 
 \cite{jain}
and the phase transitions between plateaus \cite{zirn}.
The non-commutative Chern-Simons theory, if appropriate, 
could tackle these problems being actually non-relativistic,
while the methods of non-commutative geometry could provide new 
theoretical tools.

Susskind's proposal was analysed by Polychronakos \cite{poly1,poly3},
who introduced a finite-dimensional regularization of the 
non-commutative Chern-Simons theory, the so-called Chern-Simons
matrix model or, more precisely, matrix quantum mechanics.
This theory is suitable for describing finite systems, that 
expose the relevant boundary excitations; 
on the other hand, the original field theory of infinite fluids is 
a topological theory that is not fully defined without specifying
the ultraviolet (and infrared) regularizations.

Furthermore, Polychronakos showed that the matrix model possesses
a U(N) gauge symmetry, where N is the size of the matrices, and that it
can be reduced to 2N physical degrees of freedom living on a two-dimensional
phase space. The resulting reduced theory is the Calogero model
of one-dimensional non-relativistic fermions with repulsive interaction.
This model has many features in common with the Laughlin theory
in the lowest Landau level, but is not equivalent.
More precisely, the two quantum problems have isomorphic set of states 
but different measures of integration, that are real one-dimensional 
and complex two-dimensional, respectively.
On the other hand, the classical solutions of the non-commutative
theory present the expected features of the incompressible Hall fluid 
and its vortex excitations with fractional charge.
The expected Hall conductivity was also derived in Ref.\cite{hansson}.

Karabali and Sakita \cite{karabali} analysed the reduction of 
the matrix theory to the complex eigenvalues using the coherent 
states of the electrons in the lowest Landau (Bargmann-Fock space).
They could not disentangle the electron coordinates (the complex
eigenvalues) from the auxiliary variables of the boundary fields,
but could perform some explicit calculations for low N. 
They found that the overlaps of states contain
the Laughlin wave function together with a non trivial measure 
of integration that modifies the properties of the incompressible
fluid at short distances.
These authors concluded that either the matrix model does not
describe the Laughlin physics at all, or it does it in another, unknown
set of variables that is hard to find in general.

In this paper, we shall analyse the relation of the Chern-Simons
matrix model with the Laughlin Hall states along similar lines.
Using a canonical change of matrix variables, we
solve the Gauss law constraint and reduce the theory to the physical
degrees of freedom that are the complex eigenvalues and their canonical
conjugate momenta.
Thus, the path integral of the matrix theory becomes the
holomorphic path integral of the lowest Landau level and the
eigenvalues can be interpreted as electron coordinates.
The states expressed in terms of the complex eigenvalues display
the Laughlin wave function, while the overlap integrals are 
again different from the expected form.
The latter fact can be explained as follows.

In the reduced variables, we find that the derivatives are replaced 
by covariant derivatives, similar to those caused
by an ordinary Chern-Simons interaction that is
solved instantaneously in terms of the sources (the so-called
statistical interaction \cite{wilczek}).
Therefore, a kind of statistical interaction is present among the
electrons (yet keeping their fermion statistics).
In the Bargmann space, the derivatives correspond to the 
conjugate variables, thus the statistical interaction modifies
the rule of conjugation and leads to non-standard expressions for
the overlap integrals.

The statistical interaction is actually necessary for the stability of
the Laughlin ground state: upon acting with the
covariant derivatives, it is not possible to create an excitation
with energy and angular momentum lower than those of the Laughlin state.
Since a reduction of angular momentum amounts to a compression of
the fluid, energy stability also corresponds to incompressibility of the
quantum Hall fluid.

The incompressibility of the ground state can be described by the 
highest-weight conditions for
a representation of the (non-relativistic) $W_\infty$ algebra 
of quantum area-preserving diffeomorphisms \cite{fletcher}: 
actually, this symmetry characterizes the quantum Hall fluids 
and their excitations \cite{winf}.
We thus analyse the realization of the $W_\infty$ algebra in
the Chern-Simons matrix model, both in the original and the 
reduced variables, and prove the highest-weight conditions satisfied
by the ground state using the covariant derivatives.
However, in this paper we cannot derive the complete representation
of the $W_\infty$ algebra, owing to normal-ordering
and finite-size problems.
We remark that the state overlaps can be expressed
as commutators of the $W_\infty$ algebra, such that the $W_\infty$ 
symmetry, once fully understood, can provide a complete 
algebraic characterization of the Laughlin incompressible fluids 
that is independent of the choice of coordinates.

In conclusion, the Chern-Simons matrix model exactly describes
the Laughlin Hall states if it realizes the $W_\infty$ symmetry.
Although we cannot presently prove this, we believe that 
our results are rather positive and worth discussing.

The plan of the paper is the following: 
in section two, we recall the Susskind approach, the Polychronakos
matrix model and the relation with the Calogero model.
In section three, we discuss the holomorphic quantization of the
matrix model, solve the Gauss law and obtain the Laughlin wave function
of the complex eigenvalues. 
In section four, we discuss the 
realization of the $W_\infty$ symmetry in the matrix model,
its relation with the ground state stability and the physical
interpretation of the incompressible Hall fluids.
In section five, we perform the reduction to the eigenvalues of
the path integral and the state overlaps, and obtain
the dynamics of electrons in the lowest Landau level.
The same analysis in the case of real quantization yields the path
integral of the Calogero model.
In section six, we discuss our results and suggest some 
developments.


\section{Non-commutative Chern-Simons theory, Chern-Simons matrix model
and Calogero model}

The quantization of non-commutative field theories and especially of
the topological Chern-Simons theory presents several subtle technical
aspects, that sometimes have led to inconsistencies in the literature.
Thus, we want to describe our approach as clearly as possible,
and start with a short but self-contained introduction
to Susskind's derivation \cite{susskind} and 
the developments by Polychronakos and other authors
\cite{poly1,poly3,heller,karabali,jackiw}.

Let us begin with N first-quantized electrons with two-dimensional coordinates 
$X^a_\a(t)$, $a=1,2$, $\a=1,\dots,N$, subjected to a strong 
magnetic field $B$ such that their action can be
projected to the lowest Landau level \cite{dunne},
\be
S\ =\ \frac{eB}{2}\ \int dt\ \sum_{\a=1}^N\ \e_{ab}\ X^a_\a\ \dot{X}^b_\a
\ .
\label{S-LLL}
\ee
Susskind considered the limit of the continuous fluid \cite{susskind}:
\be
\vec{X}_\a(t)\ \to\ \vec{X}(\vec{x},t)\ ,
\qquad\qquad \vec{X}(\vec{x},t=0)\ =\ \vec{x}\ ,
\label{lag-fluid}
\ee
where $\vec{x}$ are the coordinates of an initial, reference configuration
of the fluid. 
The resulting fluid mechanics is in the Lagrangian formulation,
because the field $\vec{X}$ follows the motion of the fluid \cite{jackiw}. 
For incompressible fluids, the constraint of constant density, 
$\r(\vec{x})=\r_o$, can be written in terms of Poisson brackets 
$\{\cdot,\cdot\}$ of the $\vec{x}$ coordinate as follows:
\be
\r_o\ =\ \r(\vec{x})\ =
\r_o\left\vert\frac{\de\vec{X}}{\de\vec{x}}\right\vert \ =\ 
\frac{\r_o}{2}\ \e_{ab}\ \{X^a,X^b\}\ .
\label{incompr}
\ee
This constraint can be added to the action by using
the Lagrange multiplier $A_0$, 
\be
S\ =\ \frac{eB\r_o}{2}\ \int dt\ d^2x\ 
\left[\e_{ab}\ X^a\left(\dot X^b -\th\{X^b,A_0\}\right)
\ + \ 2\th\ A_0 \right]\ ;
\label{S-winf}
\ee
in this equation, we introduced the constant $\th$,
\be
\th \ =\ \frac{1}{2\pi\r_o}\ ,
\label{theta-def}
\ee
that will later parametrize the non-commutativity.

The action (\ref{S-winf}) is left invariant by reparametrizations 
of the $\vec{x}$
variable with unit Jacobian, the area-preserving diffeomorphism,
also called $w_\infty$ transformations \cite{fletcher}\cite{winf}: they 
correspond to changes of the original labels of the fluid at $t=0$ 
(cf. Eq.(\ref{lag-fluid})) \cite{susskind}\cite{jackiw}.
The $w_\infty$ symmetry can be put into the form of a gauge invariance 
by introducing the gauge potential $\vec{A}$, as follows:
\be
X^a\ =\ x^a +\ \th\ \e_{ab}\ A_b(x)\ .
\label{a-def}
\ee
The action (\ref{S-winf}) can be rewritten in the Chern-Simons form 
in terms of the three-dimensional gauge field $A_\mu=(A_0,A_a)$:
\be
S\ =\ -\frac{k}{4\pi}\int dt\ d^2x\ \e_{\mu\nu\r}\left(
\de_\mu A_\nu A_\r \ +\ \frac{\th}{3}\left\{A_\mu,A_\nu\right\}A_\r
\right)\ .
\label{w-CS}
\ee
The coupling constant $k$ parametrizes the filling fraction of this 
(semi)classical fluid:
\be
\nu^{(cl)}\ = \ \frac{2\p\r_o}{eB}\ = \ \frac{1}{eB\th} =\ \frac{1}{k}.
\label{nu-class}
\ee

After the analysis of Lagrangian incompressible fluids, 
Susskind made a proposal for the complete theory of the fractional
quantum Hall effect, that could hold beyond the continuous 
fluid approximation by accounting for the granularity of the electrons.
He suggested to replace the theory (\ref{w-CS}) with 
the non-commutative (Abelian) Chern-Simons theory \cite{NCFT},
\be
S_{NCCS}\ =\ -\frac{k}{4\pi}\int dt\ d^2x\ \e_{\mu\nu\r}\left(
\de_\mu A_\nu \star A_\r \ -\ \frac{2i}{3}A_\mu\star A_\nu\star A_\r
\right)\ ,
\label{S-NCCS}
\ee
where the Moyal star product is:
\be
\left(g\star f\right)(x)\ = \left.\ \exp\left(i\ \frac{\th}{2}\ \e_{ab}\ 
\frac{\de}{\de x^a_1}\frac{\de}{\de x^b_2}\right)\ 
f(x_1)\ g(x_2) \right\vert_{x_1=x_2=x} \ .
\label{moyal}
\ee
Actually, the two actions (\ref{S-NCCS}) and (\ref{w-CS}) 
agree to leading order in $\theta$, i.e. for dense fluids.
In the new action (\ref{S-NCCS}), the gauge fields with Moyal product 
have become Wigner functions of the 
non-commuting operators, $\wh{x}_1,\wh{x}_2$, the former spatial 
coordinates \cite{ncgeom}:
\be
\left[\wh{x}_1,\wh{x}_2 \right] = x_1\star x_2 -x_2\star x_1=i\ \th \ .
\label{x-comm}
\ee
The corresponding quantization of the area can be thought of as a 
discretization of the fluid (at the classical level),
with the minimal area $\th$ allocated to a single electron \cite{susskind}.
Other motivations for this proposal were found in
the study of  D-branes dynamics \cite{susskind2}.

Any non-commutative field theory corresponds to a theory of
infinite-dimensional matrices, that represent the 
commutator $\left[\wh{x}_1,\wh{x}_2 \right] = i\ \th $.
The general map can be found in Ref.\cite{NCFT}, while the specific 
case of the Chern-Simons theory has been discussed e.g. in 
Ref.\cite{susskind2}.
The matrix theory equivalent to (\ref{S-NCCS}) 
is the Chern-Simons matrix quantum mechanics (Chern-Simons matrix model)
with action:
\be
S_{CSMM}\ = \ \frac{eB}{2}\ \int dt\ \Tr \left[ 
\e_{ab}\ \wh{X}^a\left(\dot{\wh{X}}^b\ +\ i[ \wh{X}^b ,\ \wh{A}_0]
\right)\ + \ 2\th\ \wh{A}_0 \right]\ .
\label{S-CSMM}
\ee
The variation of this action with respect to $\wh{A}_0$
yields the Gauss-law constraint,
\be
[\wh{X}^1\ ,\ \wh{X}^2]\ = \ i \th\ , 
\label{Gauss-cond}
\ee
that can actually be solved in terms of infinite-dimensional
Hermitean matrices $\wh{X}^a$ (and thus  $\wh{A}_0$).
The previous relation (\ref{a-def}) between the gauge field $A_a$ 
and the coordinates $X^a$ still holds for the hatted matrix
variables: upon substituting it in the matrix action (\ref{S-CSMM}),
one recovers the non-commutative theory (\ref{S-NCCS}) \cite{susskind2}.
Let us finally note that the correspondences between the matrix 
(\ref{S-CSMM}) and original (\ref{S-NCCS}) theories. 

Susskind proposal has been analyzed by many authors that 
found several evidences of quantum Hall physics, 
both at the classical and at the quantum level.
However, there remain some open problems; two of them will be
particularly important for our discussion:

\begin{itemize}
\item In the extension from the fluid (\ref{S-winf}) to the non-commutative
theory (\ref{S-CSMM}), the electron coordinates $\vec{X}$ become 
matrices and loose their physical interpretation. 
There is the question of defining the physical
(gauge-invariant) observables in the matrix theory that correspond to
the electron coordinates, the density $\r(\vec{x})$ and other quantities.
This issue has been addressed in the Refs.\cite{karabali}\cite{hansson}.

\item The Gauss law (\ref{Gauss-cond}) admits one solution 
modulo reparametrizations, therefore the matrix theory (\ref{S-CSMM}) 
possesses just one state, i.e. the ground state of an infinitely extended 
incompressible fluid with infinite electrons.
This is a topological theory with rather peculiar quantum properties:
for example,  the number and type of physical degrees of freedom
may depend on the regularization and the boundary conditions;
they should suitably chosen for the complete definition of the theory.
\end{itemize}
An answer to the second question was given by Polychronakos \cite{poly1} who
regularized the matrix model by introducing a
``boundary'' vector field $\psi^i$, $i=1,\dots,N$, 
such that the modified Gauss law admits finite-dimensional matrix solutions.
The modified action is:
\ba
S_{CSMM} & = & \frac{B}{2}\ \int dt\ \Tr \left[ 
\e_{ab}\ X^a\left(\dot{X}^b\ -\ i[ A_0\ ,\ X^b]
\right)\ + \ 2\th\ A_0 -\w\ (X^a)^2\right] \nl
&& -\ \int dt\ \psi^\dag \left(i\dot{\psi} \ +\ A_0 \psi \right)\ .
\label{S-Poly}
\ea
In this equation, we suppressed the hats over the matrices, set
$e=1$ and also introduced a quadratic potential with coupling $\w$.
The Gauss law now reads:
\be
G \ =\ 0\ , \qquad \qquad 
G\ =\ -i B\ \left[X_1,X_2\right]\ -\ B\th\ +\ \psi \psi^\dag\ .
\label{gauss-Poly}
\ee
The condition $\Tr\ G = 0$ can be satisfied by $N\times N$-dimensional
matrices $X^a$, provided that $\Vert\psi\Vert^2=NB\th$. 
This reduction to a finite-dimensional quantum mechanical problem
introduces both an ultraviolet and an infrared cutoff in the theory.
As shown in Ref.\cite{poly1}, one can choose a gauge in
which $\psi^i$ has only one non-vanishing component, say the N-th one,
resulting into a boundary effect that disappears for large N
upon defining a suitable weak limit $N\to\infty$.
The $U(N)$ symmetry of the matrix Chern-Simons theory (\ref{S-Poly})
is given by $X^a \to U X^a U^\dag$
and $\psi \to U\psi$, where $U$ is a unitary matrix.
Since $G$ is the generator of infinitesimal transformations at the
quantum level, the Gauss-law condition requires that all
physical states should be $U(N)$ singlets \cite{poly1,susskind}.

Classical matrix solutions of this theory were found in Ref.\cite{poly1}
for the ground state and the quasi-hole excitation that possess
the expected features of Hall incompressible fluids.
The distribution of the matrix eigenvalues for 
the ground state in the isotropic potential $\w \Tr \vec{X}^2$ 
is a circular droplet with uniform density $\r_o\sim 1/2\pi\th$
for large N, as in the infinite theory.
In the quasi-hole solution, a hole (vortex) is present
in the density with the correct size.
These solutions suggest the identification of the matrix eigenvalues
with the electron coordinates: however, only one
matrix can be diagonalized, say $(X^1)_{nm}= x^1_n \d_{nm}$,
and their distribution $\r(x^1)$ is actually meant to be integrated over
the other coordinate $x^2$.


\subsection{Quantization on the real line}

The quantization of Polychronakos' theory (\ref{S-Poly}) can be done before or
after having solved the Gauss constraint: both approaches have been considered
in Ref.\cite{poly1} and we discuss the latter first.
Since the Hermitean matrix $X^1$ can be diagonalized 
by a unitary transformation, we can fix the gauge:
\be
(X^1)_{nm}\ =\ x_n\ \d_{nm} \ .
\label{X-diag}
\ee
The form of the other variables is obtained by solving the Gauss 
constraint (\ref{gauss-Poly}) in this gauge. The result is:
\ba
\psi_n &=& \sqrt{B\th}\ ,\qquad \forall \ n\ ,\nl
(X^2)_{nm} &=& y_n \d_{nm} \ -\ i\th\ \frac{1-\d_{nm}}{x_n-x_m} \ ;
\label{sol-gauss}
\ea
these two equations follow from the diagonal and off-diagonal 
components of $G_{nm}=0$, respectively; in the diagonal components,
we used the residual $U(1)^N$ gauge symmetry to fix the phases
of $\psi_n$.
The variables $y_n$ in (\ref{sol-gauss}) parametrize the components of $X_2$ 
that are left undetermined.
The substitution of all variables back into the action yields:
\ba
S &=& - \int dt\ B \ \sum_{n=1}^N\ \dot{x}_n\ y_n \ +\ H \ ,\nl
H &=& \frac{B\w}{2}\ \Tr\ \vec{X}^2\ =\ \sum_{n=1}^N \left(
\frac{\w}{B}\frac{p^2_n}{2} \ +\ \frac{B\w}{2} x_n^2 \right)\ +\ 
\sum_{n\neq m}^N\ \frac{\w B\th^2}{2}\ \frac{1}{(x_n-x_m)^2} \ .
\label{S-Calog}
\ea
Namely, the interpretation of the N real variables $x_n$ as particle
coordinates has led to the identification of the
conjugate momenta $p_n=- By_n$. 
Moreover, the Hamiltonian is found to be that of the Calogero
model with coupling constant $B\th=k$ taking integer values\footnote{
The quantization of the Chern-Simons coupling $k$ 
follows from the requirement of invariance of the action (\ref{S-NCCS})
under large gauge transformation \cite{nair}.}.
Therefore, the Chern-Simons matrix model has been reduced
to the quantum mechanics of $N$ particles on the line with 
two-body repulsion. Note the reduction of degrees of freedom: 
starting from the $2N^2 +2N$ real phase-space variables 
$(X_1,X_2,\psi,\psi^\dag)$, the gauge fixings eliminate
$N^2$ variables and the Hermitean constraint $G\equiv 0$ further 
$N^2$ ones, leaving the conjugate variables of $N$ particles.

The one-dimensional Calogero model is closely related to the theory
of two-dimensional electrons quantized in the first Laudau level:
it is integrable and the space of states is known \cite{brink}
and isomorphic to that of the excitations over the Laughlin state 
at filling fraction $\nu=1/(k+1)$ \cite{laugh};
the Calogero particles satisfy selection rules of 
an enhanced exclusion principle \cite{poly5} that allow to define a 
one-dimensional analog of the fractional statistics of anyons
\cite{wilczek}.
On the other hand, the Hilbert spaces of the two problems are
different, because the one-dimensional norm of the Calogero
model is different from that of the first
Landau level \cite{brink,pasquier}.

Therefore, Polychronakos' analysis found strong analogies
between the Chern-Simons matrix model and the Laughlin Hall states
but not a complete equivalence. 
In the next section, we shell discuss another quantization scheme 
and will perform a change of matrix variables that let
the electron coordinates and the Laughlin wave function emerge
rather naturally.


\section{Holomorphic quantization of the Chern-Simons matrix model}

We now discuss the ``covariant quantization'' of the Chern-Simons 
matrix model (\ref{S-Poly}): we solve the  
Gauss-law constraint at the quantum level, elaborating on the 
results of the Refs. \cite{poly1} \cite{heller}.
It is convenient to introduce the complex matrices:
\be
X\ =\ X_1\ +\ i\ X_2\ ,\qquad\qquad X^\dag\ =\ X_1\ -\ i\ X_2\ .
\label{X-mat}
\ee 
The matrix action (\ref{S-Poly}) in the $A_0=0$ gauge is:
\ba
\left. S_{CSMM} \right\vert_{A_0=0} &=& \int
dt\ \frac{B}{2}\ \Tr \left( X_1\ \dot{X}_2 \ -\ \dot{X}_1\ X_2 \right)\ 
- \ i \psi^\dag \ \dot{\psi} \ -\ {\cal H}(X_a)\nl
&=& \int dt\ \frac{B}{2i}\ \sum_{nm}\ \dot{X}_{nm}\ \ov{X}_{nm}\ 
-i\ \sum_n \dot{\psi}_n\ \ov{\psi}_n \ -\ {\cal H}(X^\dag X)\ ,
\label{S-complex}
\ea 
where ${\cal H}=B\w\ \Tr\ X_a^2/2$.
This action implies the following oscillator commutation relations 
between the components of the matrices and vectors:
\ba
\left[\left[\ \ov{X}_{nm}\ ,\ X_{kl} \ \right]\right] & = &
\frac{2}{B}\ \d_{nk}\ \d_{ml}\ ,\nl
\left[\left[\ \ov{\psi}_{n}\ ,\ \psi_{m}\ \right]\right] & = &
\ \d_{nm}\ . 
\label{X-comm}
\ea 
In these equations, we represented the quantum commutator with
double brackets to distinguish it from the classical matrix 
commutator.

The form of the action (\ref{S-complex}) is that of $N^2+N$ ``particles''
in the lowest Landau level with complex ``coordinates'' $X_{nm}$ and $\psi_n$, 
that can be quantized in the Bargmann-Fock space
of holomorphic wave functions $\Psi(X,\psi)$, 
with integration measure \cite{faddeev}: 
\be
\left\la\Psi_1 \vert \Psi_2 \right\ra\ = \ 
\int {\cal D}X\ {\cal D}\ov{X}\ {\cal D}\psi\ {\cal D}\ov{\psi}\ \ 
e^{-\frac{B}{2}\Tr X^\dag X -\psi^\dag \psi }\ \ 
\ov{\Psi_1(X,\psi)}\ \Psi_2(X,\psi)\ .
\label{Fock-meas}
\ee 
The conjugate variables act as derivative operators on the wave functions,
\be
\ov{X}_{nm} \ \to\ \frac{2}{B}\frac{\de\ \ }{\de X_{nm}}\ , \qquad\qquad
\ov{\psi}_n \ \to\ \frac{\de\ }{\de \psi_n}\ ,
\label{der-def}
\ee 
and the (properly normal-ordered) Gauss law (\ref{gauss-Poly}) 
becomes a differential equation for the
wave functions of physical states ($B\th=k$):
\ba
&& G_{ij}\ \Psi_{\rm phys}(X,\psi) \ =\  0\ , \nl
&& G_{ij}\ =\ \sum_\ell \left( X_{i\ell}\ \frac{\de\ \ }{\de X_{j\ell}} \ -\ 
 X_{\ell j}\ \frac{\de\ \ }{\de X_{\ell i}} \right) \ -\
k\ \d_{ij}\ +\ \psi_i\ \frac{\de\ }{\de \psi_j} \ . 
\label{diff-gauss}
\ea 

We now come to a crucial point of our analysis: we are going to
perform a change of matrix variables that leaves invariant 
the commutation relations (\ref{X-comm}). 
This B\"acklund (or Bogoliubov) transformation is defined as follows:
\ba
X &=& V^{-1}\ \L\ V\ ,\qquad\qquad \L\ = \ diag(\l_1,\dots,\l_N)\ , \nl
\psi &=& V^{-1}\ \phi\ .
\label{Bogo-trans}
\ea
Here we used the fact that a complex matrix can be diagonalized
by a $GL(N,\C)$ transformation (up to the zero-measure set of matrices with
degenerate eigenvalues).
For the transformation of the derivative operators (\ref{der-def}),
we should consider the linear transformation in the $N^2+N$ 
dimensional tangent space: 
\be
\left\{dX_{nm}\ ,\ d\psi_n\right\}\ \to\ 
\left\{d\l_n \ ,\ dv_{ij}\ (i\neq j)\ ,\ d\phi_n \right\}\ , 
\qquad \qquad dv\ = \ dV\ V^{-1}\ . 
\label{tang-trans}
\ee 
From the transformation of the covariant vector  
$\left\{dX_{nm}\ ,\ d\psi_n\right\}$,
\ba
dX &=& V^{-1}\left( d\L \ +\ [\L, dv]\right) V\ ,\nl
d\psi &=& V^{-1}\left(d\phi \ -\ dv\ \phi \right)\ ,
\label{d-trans}
\ea
we can compute the inverse transformation of the 
contravariant vector
$\left\{\de/\de X_{nm}\ ,\ \de/\de\psi_n\right\}$.
The result is the following:
\ba
\frac{\de\ \ }{\de X_{ij}} &=&
V_{ni}\ V^{-1}_{jm}\ \frac{\de\ \ }{\de \L_{nm}}\ , \qquad
\frac{\de\ \ }{\de \L_{nm}} \ =\ 
\frac{\de\ }{\de\l_n}\ \d_{nm}\ +\ \frac{1-\d_{nm}}{\l_n-\l_m}\ 
\left(\frac{\de\ \ }{\de v_{nm}}\ +\ 
\phi_m\ \frac{\de\ }{\de\phi_n}\right)\ ,
\nl
\frac{\de\ }{\de \psi_j} &=& V_{nj}\ \frac{\de\ }{\de \phi_n} \ .
\label{der-trans}
\ea 
This transformation is invertible for 
$\det V \neq 0$ and distinct eigenvalues $\l_n\neq\l_m$.
One can explicitly check that the commutation relations (\ref{X-comm})
are left invariant by the transformation, namely that
each new variable $(\l,V, \phi)$
satisfies canonical commutators with the corresponding 
derivative\footnote{
The transformation of the matrix derivative has been suitably normal-ordered 
in Eq.(\ref{der-trans}).}.

The oscillator vacuum state is left invariant by 
the transformation (\ref{Bogo-trans}): indeed,
the vacuum wave functions for the original 
and new oscillators, 
$\Psi^{\rm (old)}_o=1$ and $\Psi^{\rm (new)}_o$, respectively,
should satisfy:
\ba
\frac{\de\ \ }{\de X_{ij}}\ \Psi^{\rm (old)}_o & =& 0\ ,\qquad\qquad
\frac{\de\ }{\de \psi_j} \ \Psi^{\rm (old)}_o\ =\ 0\ , \nl
\frac{\de\ \ }{\de v_{ij}}\ \Psi^{\rm (new)}_o & =& 0\ ,\ i\neq j,\qquad
\frac{\de\ }{\de \l_j} \ \Psi^{\rm (new)}_o\ =\ 
\frac{\de\ }{\de \phi_j} \ \Psi^{\rm (new)}_o\ =\ 0\ .
\label{vac-cond}
\ea
The comparison of these expressions with the transformation 
(\ref{der-trans}) shows that it is consistent to keep the same vacuum:
$\Psi^{\rm (new)}_o=1$.
This result is at variance with the usual Bogoliubov transformations, 
where the new vacuum contains an infinite number of old particles.
Therefore, the transformation (\ref{der-trans}) preserves the number operators
associated to both $X_{ij}$ and $\psi_i$ oscillators.

The substitution of the new matrix variables 
(\ref{Bogo-trans}) and derivatives 
(\ref{der-trans}) in the Gauss law (\ref{diff-gauss})
yields the following result:
\ba
G_{ij} &=& V^{-1}_{im}\ V_{nj}\ \wt{G}_{nm}\ ,\qquad\qquad 
\wt{G}_{nm}\ \Psi(\L,V,\phi)\ = \ 0\ ,\nl
\wt{G}_{nm} &=& \left\{
\begin{array}{ll}
- \frac{\de\ \ }{\de v_{nm}} & n\neq m\ ,\\ & \\
\phi_n\ \frac{\de\ }{\de \phi_n}\ -\ k& n = m\ .
\end{array}
\right.
\label{gauss-diag}
\ea
Rather remarkably, the change of variables 
diagonalizes the constraint  and allows for the 
elimination of the unphysical degrees of freedom:
\begin{itemize}
\item The $N^2-N$ off-diagonal components of $V$ are killed,
namely $\Psi^{\rm (phys)}(\l,V,\phi)$ can only depend on $V$
through quantities like $\det V$.
\item  The $N$ degrees of freedom of $\psi$ are also frozen,
because all physical wave functions should contain the same
homogeneous polynomial of degree $k$ in each component of the
vector, which is $\prod_{n=1}^N\ (\phi_n)^k$.
\end{itemize}

The remaining dynamical variables are 
the $N$ complex eigenvalues $\l_n$, that can be interpreted as
coordinates of electrons in lowest Landau level.


\subsection{Wave functions}

The general solution of the Gauss law for the wave functions of
physical states in the $(X,\psi)$ coordinates,
 Eq. (\ref{diff-gauss}), has been found in the 
Refs. \cite{poly1} \cite{heller}: we should form
$U(N)$-singlet polynomials made of
the $N$-component epsilon tensor and an arbitrary number of 
$X$ matrices; moreover, the condition $\Tr\ G=0$ in 
(\ref{diff-gauss}) implies that the vector $\psi$ 
should occur to the power $Nk$.
For $k=1$, these wave functions take the form \cite{heller}:
\be
\Psi_{\{n_1,\dots,n_N\}}\left(X,\psi\right)\ = \ 
\eps^{i_1\dots i_N}\ \left(X^{n_1}\psi\right)_{i_1}\cdots
\left(X^{n_N}\psi\right)_{i_N} \ ,\qquad 0\le n_1<n_2<\cdots <n_N\ .
\label{HVR-states}
\ee
for any ordered set of positive integers $\{n_i\}$.
The ground state in the confining potential $\Tr( X X^\dag)$
is given by the closest packing $\{0,1,\dots,N-1\}$ that has the lowest
degree in $X$.
For $k\neq 1$, one can multiply $k$ terms of this sort, leading to
$\Psi_{\{n^1_1,\dots,n^1_N\}\cdots\{n^k_1,\dots,n^k_N\}}$.
As shown in Ref.\cite{heller}, there is an equivalent
basis for these states that involves the ``bosonic'' powers
of $X$:
\ba
\Psi \left(X,\psi\right) &=& \sum_{\{m_k\}}\ \Tr \left(X^{m_1}\right)\cdots
\Tr \left(X^{m_k}\right) \ \Psi_{k-gs}\ ,\nl
\Psi_{k-gs} &=& \left[\eps^{i_1\dots i_N} \ 
\psi_{i_1}\ \left(X\psi\right)_{i_2}\cdots \left(X^{N-1}\psi\right)_{i_N}
\right]^k\ ,
\label{HVR-bose}
\ea
where the positive integers $\{m_1,\dots,m_k\}$ are now
unrestricted.
This second basis (\ref{HVR-bose}) also makes sense in the
$k=0$ case, where $\Psi_{0-gs}=1$.

Let us now perform the change of matrix variables in the wave functions
(\ref{HVR-bose}): the excitations made by the invariant powers 
$\Tr (X^r)$ became the power sums of the eigenvalues, $\sum_n \l^r_n$; 
in the ground state wave function, the
dependence on $V$ and $\phi$ factorizes and the powers of the eigenvalues
make up the Vandermonde determinant $\D(\l)=\prod_{i<j}(\l_i-\l_j)$: 
\ba
\Psi_{k-gs}\left(\L,V,\psi\right) &=& \left[ \eps^{i_1\dots i_N} \ 
\left(V^{-1}\phi\right)_{i_1}\ 
\left(V^{-1}\L\phi\right)_{i_2}\cdots 
\left(V^{-1}\L^{N-1}\phi\right)_{i_N} \right]^k\ \nl
&=&\left[\left(\det V\right)^{-1} \ 
\det \left( \l_j^{i-1}\ \phi_j \right) \right]^k \nl
&=& \left(\det V\right)^{-k} 
\ \prod_{1\le n< m\le N}\left(\l_n-\l_m\right)^k\  
\left(\prod_i \ \phi_i\right)^k\ .
\label{Laugh-wf}
\ea
We thus obtain the Laughlin wave function for the ground state of
the Hall effect with the electron coordinates corresponding to the complex
eigenvalues of $X$. The value of the filling fraction is:
\be
\nu\ =\ \frac{1}{k+1}\ ,
\label{nu}
\ee
and is renormalized from the classical value (\ref{nu-class})
because the wave function should acquire
one extra factor of $\D(\l)$ from the integration measure, as shown later. 
The factorized dependence on $V$ and $\psi$ in (\ref{Laugh-wf}) 
is the same for all the states (\ref{HVR-bose}), since it is the
unique solution to the Gauss law (\ref{gauss-diag}); namely,
these degrees of freedom are frozen.
 The bosonic power sums $\sum_n \l^r_n$ are
the natural basis of symmetric polynomials forming the excitations
over the Laughlin state\footnote{
The counting of these states is given by the number of partitions
and it shows the correspondence between the edge excitations of
incompressible Hall fluids and the states of a one-dimensional 
bosonic field \cite{wen}.}.

Therefore, we have shown that the change of matrix variables (\ref{Bogo-trans})
allows to explicitly eliminate the gauge degrees of freedom and
reduce Chern-Simons matrix model to the quantum mechanics of
N variables with ground state given by the Laughlin wave function.
In the next section, we discuss the meaning of the
covariant derivatives (\ref{der-trans}) that have emerged 
in the reduction.


\section{Stability, incompressibility and $W_\infty$ symmetry}

\subsection{Introduction}

In a series of papers \cite{winf,cdtz,flohr,minmod}, 
the incompressible Hall fluids
have been characterized by the symmetry under $W_\infty$ transformations,
that are the quantization of the $\w_\infty$ area-preserving diffeomorphisms
of the plane \cite{fletcher}. 
Actually, the deformations of a classical droplet 
of fluid of constant density have all the same area and can be
mapped one into another by $\w_\infty$ reparametrizations.
In the quantum theory of the first Landau level, the $\nu=1$ 
ground state is a circular droplet of quantum incompressible fluid 
that admits the further interpretation of a filled Fermi sea \cite{winf}:
the electrons occupy all the available one-particle 
states of angular momentum
$J=0,1,\dots,N-1$, leading to a droplet of radius $R\sim \sqrt{N}$.

The deformations of the droplet are obtained from the $W_\infty$
operators: they are the moments of the generating function of
classical $\w_\infty$ transformations that are
quantized in the Bargmann space \cite{winf}:
\be
{\cal L}_{nm}\ = \ \sum_{\a=1}^N\ \l_\a^n\ 
\left(\frac{\de}{\de\l_\a}\right)^m\ ,\qquad \qquad \a=1,\dots,N,
\label{Lnm-def}
\ee
where $\a=1,\dots,N$ is the particle index,
$\bar{\l}\to\de/\de\l$ when acting on holomorphic wave functions,
and $n,m$ are non-negative integers. 

The $\winf$ operators generate small fluctuations of the $\nu=1$ ground
state\footnote{
We use the notation $\Phi$ for the wave function of the physical electron
variables, to distinguish it from that of the matrix model
$\Psi$ (\ref{Laugh-wf}).}, 
$\Phi_{gs}=\D(\l)$, that possess angular momentum $\D J= n-m$ with
respect to the ground state.
Excitations with $\D J<0$ are forbidden in the filled Fermi sea, because
they would correspond to violations of the exclusion principle.
Therefore, the ground state should satisfy the conditions:
\be
{\cal L}_{nm} \ \Phi_{gs}\ =\ 0\ , \qquad \qquad 0\le n<m\le N-1\ . 
\label{W-inc}
\ee
Furthermore, the generators with $m\ge N$ also vanish
because they would correspond to particle-hole transitions 
outside the Fermi sea \cite{winf}.

The $N(N-1)/2$ conditions (\ref{W-inc}) express
the incompressibility of the $\nu=1$ quantum Hall ground state and
represent the highest-weight conditions for the representation
of the $W_\infty$ algebra.
In the quadratic confining potential, 
${\cal H}=\sum_\a \bar{\l}_\a \l_\a \propto J$, the incompressibility
of the ground state is equivalent to its energy stability.

The other states in the infinite-dimensional representations
are the excitation obtained by acting with ${\cal L}_{nm}$, 
$n>m$, on the ground state. 
One can show that these operators generate all the bosonic power
sums $ \sum_\a \l_\a^k$
described before, such that the ${\cal L}_{nm}$ carry over
the bosonization of the incompressible fluid in the non-relativistic
theory\footnote{
Actually, these operators become the bosonic current and its
normal-ordered powers when evaluated in the relativistic 
effective theory of edge excitations \cite{cdtz}.}.

The $W_\infty$ algebra is \cite{winf}:
\be
\left[{\cal L}_{nm}\ ,\ {\cal L}_{kl} \right]
\ = \ \sum_{s=1}^{Min(m,k)}\ \frac{m!\ k!}{(m-s)!\ (k-s)!\ s!}\ 
{\cal L}_{n+k-s,\ m+l-s}\ 
-\ \left(m \leftrightarrow l\ ,\ n \leftrightarrow k\right)\ .
\label{W-alg}
\ee
The first term in the r.h.s, 
$\left[{\cal L}_{nm}\ ,\ {\cal L}_{kl} \right]=\hbar
(mk-nl){\cal L}_{n+k-1,\ m+l-1}$, corresponds to the quantization of
the classical algebra $w_\infty$
of area-preserving diffeomorphism, while the other
terms are quantum corrections $O(\hbar^p)$, $p\ge 2$.
Finally, the operator with equal indices, ${\cal L}_{nn}$, are
the Casimirs of the representation, e.g. ${\cal L}_{00}=N$,
${\cal L}_{11}=J$. The representations of the $W_\infty$
symmetry for N electrons can be related to the representation
of the $U(N)$ algebra \cite{fletcher}.

The $W_\infty$ algebra is also useful for expressing the overlaps of states.
Consider two excitations, e.g.
${\cal L}_{mn}\Phi_{gs}$ and $ {\cal L}_{kl}\Phi_{gs}$,
for $m>n$ and $k>l$: thanks to the conjugation rule,
\be
{\cal L}_{nm}^\dag\ = \ {\cal L}_{mn}\ ,
\label{W-dag}
\ee
and the incompressibility conditions (\ref{W-inc}), 
the overlap of these two states can be rewritten as a commutator:
\be
\la{\cal L}_{mn}\Phi_{gs}\vert {\cal L}_{kl}\Phi_{gs}\ra\ =\ 
\la\Phi_{gs}\vert\left[{\cal L}_{nm},{\cal L}_{kl}\right]\vert
\Phi_{gs}\ra\ , \qquad\qquad m>n,\ k>l,
\label{W-over}
\ee
that can be reduced to the Casimirs of the algebra, if non-vanishing.
In particular, one finds that  the $\nu=1$ representation is unitarity,
thanks to the positivity of the Casimirs \cite{winf}.

In conclusion, the use of the $W_\infty$ symmetry allows a complete 
algebraic description of the  $\nu=1$ incompressible quantum Hall fluid
and its excitations, that does not rely on the coordinate 
representation of wave functions and overlaps.

Let us now review previous analyses of the $W_\infty$
symmetry of the fractional Laughlin states:
\be
\Phi_{k-gs}\ =\ \D(\l)^{k+1}\ , \qquad \qquad \nu=\frac{1}{k+1}\ .
\label{red-wf}
\ee
As proposed in Ref. \cite{flohr}, one can perform a similarity transformation
on the $\nu=1$ generators, as follows:
\be
{\cal L}_{nm}^{(k)}\ =\ \D(\l)^k\ {\cal L}_{nm}\ \D(\l)^{-k}\ 
=\ \sum_{\a=1}^N\ \l_\a^n\ \left(\frac{\de}{\de\l_\a} \ -
\ \sum_{\b,\b\neq\a}\frac{k}{\l_\a-\l_\b}\right)^m\ .
\label{W-gen-nu}
\ee
These operators are non-singular when acting of the Laughlin wave
function and its excitation, obey the same algebra (\ref{W-alg}) and realize
exactly the same $\nu=1$ representation (same values of the Casimirs);
in particular, the incompressibility conditions read again:
\be
{\cal L}_{nm}^{(k)}\ \ \Phi_{k-gs}\ =\ 0\ ,\qquad\qquad 0\le n<m\le N-1\ .
\label{W-inc-nu}
\ee
One problem of these operators is the Hermiticity relation (\ref{W-dag}), 
that is not manifestly satisfied and thus the unitarity of 
the representation is not guaranteed
(unless an exotic measure of integration is introduced \cite{flohr}).
Nevertheless, a couple of remarks are suggested by this analysis:
\begin{itemize}
\item 
The similarity between Laughlin states of different $\nu$ values 
and the bosonization of the corresponding
relativistic theory on the edge \cite{cdtz} 
indicate that all Laughlin states should realize 
representations of the $W_\infty$ algebra (\ref{W-alg}).
\item
In the fractional case, the suggested $W_\infty$ generators 
(\ref{W-gen-nu}) contain covariant derivatives:
\be
D_z \ =\ \de_z \ +\ A_z\ ,\qquad A_z \ =
\ -\sum_\b\frac{k}{z-\l_\b}\ ,
\label{D-cov}
\ee 
that assign a magnetic charge of $k$ fluxes to each electron,
the excess magnetic field being ${\cal B}=\left[D_z ,\ D_{\bar z}\right]=$
\hbox{$k\ \pi\ \sum_\b\ \d^2 (z-\l_\b)$},
($A_{\bar z}=0$). 
Therefore, the covariant derivatives introduce a ``statistical
interaction'' among the electrons \cite{wilczek}, 
given by the Aharonov-Bohm phases
between electric and magnetic charges; note, however,
that this interaction does not change the statistics of electrons
for even integer values of $k$ \cite{fradkin}. 
\end{itemize}


\subsection{ $W_\infty$ symmetry of the Chern-Simons matrix model}

In the Chern-Simons matrix model, we can introduce two types of
polynomial generators that generalize (\ref{Lnm-def}) and are gauge
invariant:
\ba
{\cal L}_{nm}\ =\ \Tr\left(X^n\ X^{\dag m} \right)\ ,\nl
{\cal P}_{nm}\ =\ \psi^\dag\ X^n\ X^{\dag m}\ \psi\ .
\label{W-mat}
\ea
The second operators can be considered as finite-N corrections
to the first ones, because they involve the boundary vectors.

Both families of operators satisfy the incompressibility
conditions on the matrix ground states (\ref{HVR-bose}) for all $k$:
\ba
{\cal L}_{nm}\ \Psi_{k-gs} &=& 0\ ,\qquad\qquad 0\le n<m \ ,\nl
{\cal P}_{nm}\ \Psi_{k-gs} &=& 0\ ,\qquad\qquad 0\le n<m \ .
\label{W-inc-mat}
\ea

The proof is easily obtained in graphical form. Represent
the matrices $X_{ij}$  as oriented links, the vectors $\psi_i$ by dots,
the epsilon tensor as the N-branching root of a tree, and attach
the extrema according to the summations of matrix indices;
then, the $k=1$ wave function $\Psi_{1-gs}$ is represented 
by a tree with N branches of different lengths ranging from zero to N-1.
Note that their total length $N(N-1)/2$ is the minimal one for having a
non-vanishing expression, owing to presence of the epsilon tensor.
The ${\cal L}_{nm}$ operators, e.g. ,
\be
{\cal L}_{12}\ =\ X_{ij}\ \frac{\de}{\de X_{ik}}\ \frac{\de}{\de X_{kj}}\ , 
\label{L12}
\ee 
act on $\Psi_{1-gs}$ as follows: the derivative $\de/\de X_{ij}$ 
remove one link in a branch of the tree and identifies the 
indices $(i,j)$ at the free extrema.
Then, the matrices $X^n_{ij}$ rejoin the segments and form new branches
of different lengths or nucleate closed rings;
after this cut and paste, the tree is reformed. 
Under the action of ${\cal L}_{nm}$ with
$n<m$, the total length of the branches is lower than that of
the ground state, thus the expression vanishes.
This proves the first of Eq. (\ref{W-mat}) for $k=1$.
For general $k$, the wave function $\Psi_{k-gs}$ contains 
$k$ independent trees.
The action of  ${\cal L}_{nm}$  can cut and paste branches
of different trees, but trees cannot be joined because the orientation 
of the lines would be violated. Thus, independent trees are
reformed and the previous length counting applies again.

The action of the ${\cal P}_{nm}$ operators, e.g.,
\be
{\cal P}_{01}\ =\ \psi_i\ \frac{\de}{\de X_{ij}}\ \frac{\de}{\de\psi_j}\ , 
\label{P12}
\ee
is analogous, with the addition that branches can be cut
and joined at their end points, and terminated at some point.
The resulting tree is similarly shortened for $n<m$.

Therefore, the incompressibility conditions (\ref{W-inc-mat}) are verified.
It is interesting to note the existence of
a generalized exclusion principle
in the ground state, that actually follows from the $SU(N)$ singlet condition. 
The different branches of the tree can be associated to ``states'' and 
there cannot be more than $k$ ``particles'' of the same type in 
$\Psi_{k-gs}$. The $\winf$ generators map branches into
branches, i.e. make particle-hole transitions as in the $\nu=1$ 
filled Fermi sea, some of which are forbidden by the 
close packing conditions\footnote{
The use of $SU(N)$ singlets for building states with exclusion statistics
was also proposed in Ref.\cite{read}.}.

We now discuss the algebra of two ${\cal L}_{nm}$ operators.
Some care should be taken in dealing with objects that are
both operator and matrix ordered: in fact, one should abandon
the convention of implicitly summing over contiguous matrix 
indices and leave them explicit.
The ${\cal L}_{nm}$ commutator reads (the graphical representation
is still useful):
\be
\left[\left[{\cal L}_{nm}\ ,\ {\cal L}_{kl} \right]\right] \ =\ 
X^n_{ij}\ \left[\left[\left(\frac{\de}{\de X}\right)^m_{ij}\ ,\ 
X^k_{pq}\right]\right]\ \left(\frac{\de}{\de X}\right)^l_{pq} 
\ - \ \left( {n \leftrightarrow k,\ m \leftrightarrow l \atop
i \leftrightarrow p,\ j \leftrightarrow q} \right)\ .
\label{W-alg-mat}
\ee
Again one derivative kills one matrix and identifies its pair of
indices: the results is an operator containing $(n+k-s)$ times $X$ and
$(m+l-s)$ times $\de/\de X$, with $s=1,2,\dots$.
Operator orderings are dealt with as in the case of the $\nu=1$
$\winf$ algebra (\ref{W-alg}) and they create further terms with $s\ge 1$.
However, the matrix summations in the resulting operators
may not be properly ordered for identifying them as
${\cal L}_{n+k-s,m+l-s}$; an example is,  
$ X^p_{ij}\ X^q_{kl}\ X^\dag_{jk}\ X^\dag_{li}$. 
Here we can use the Gauss law to perform the matrix reorderings because it
is an identity in gauge invariant expressions: from Eq.
(\ref{diff-gauss}), we read that the reordering of the pair 
$X_{lj}\ X^\dag_{il}\ \to\ X_{il}\ X^\dag_{lj}$
creates the extra terms $k \d_{ij}$ and $\psi_i \psi^\dag_j$,
leading to the descendants ${\cal L}_{n+k-s,m+l-s}$ and
the operators ${\cal P}_{n+k-s,m+l-s}$, with $s>1$.

Therefore, the r.h.s. of the ${\cal L}_{nm}$ algebra 
(\ref{W-alg-mat}) contains the characteristic leading $O(\hbar)$ 
term for the semiclassical interpretation, but also 
involves the finite-N descendants ${\cal P}_{n-s,m-s}$,
i.e. it does not close.
Presumably, for $N\to\infty$ the latter terms can be disregarded and
the algebra closes as in the $\nu=1$ case (\ref{W-alg}), up to possible
redefinitions of the higher, non-classical structure 
constants.
For finite N, there remain the open problem of selecting the
right basis,
\be
\wt{\cal L}_{nm}\ =\ {\cal L}_{nm} \ +\ \g_1\ {\cal P}_{n-1,m-1}\ +
\ \g_2\ {\cal P}_{n-2,m-2}\ + \ \cdots\ ,
\label{W-red}
\ee
that form a closed algebra. The incompressibility conditions
(\ref{W-inc-mat}) are satisfied anyhow.
This solution of the matrix ordering is also important for identifying 
physical quantities like the one-particle density, that is 
the generating function of $ \wt{\cal L}_{nm}$ \cite{hansson}.
 
In conclusion, we have shown that the Chern-Simons matrix model 
realizes highest-weight representations of the type
occurring in the quantum Hall effect, but we cannot presently account
for the complete form of the $\winf$ algebra.


\subsection{$\winf$ symmetry in physical coordinates}
 
The realization of the symmetry
in the electron coordinates is more interesting because it has direct
physical interpretation in the quantum Hall effect.
The form of the $\winf$ generators is obtained by replacing
the canonical transformations (\ref{Bogo-trans},\ref{der-trans}) into
the matrix expressions (\ref{W-mat}):
\ba
{\cal L}^{(k)}_{nm}& =& 
\sum_{j=1}^N\ \l^n_j\ D^m_{jj}\ ,\qquad\qquad
D_{pq}\ =\ \d_{pq}\ \frac{\de}{\de\l_p}\ -\ 
\frac{1-\d_{pq}}{\l_p-\l_q}\ \phi_p\ \frac{\de}{\de\phi_q}\ ,\nl
{\cal P}^{(k)}_{nm}& =& \sum_{i,j=1}^N\ 
\phi_i\ \l^n_i\ D^m_{ij}\ \frac{\de}{\de\phi_j}\ .
\label{W-exp}
\ea
The matrix covariant derivative $D_{ij}$ 
enforces a kind of statistical interaction among the electrons
similar to the one discussed in section 4.1. 
The form of $D_{ij}$ is obtained from (\ref{der-trans}), 
by suppressing the $V$ dependence absent in physical states,
but leaving the $\phi_i$ derivatives to allow for possible normal orderings:
one should eventually replace $\phi_i\ \de/\de \phi_i \to k$,
$\forall i$.
In writing the expressions (\ref{W-exp}), we neglected further
operator ordering problems, that are not well defined anyhow
for non-linear transformations. 
Therefore, the present operators are not guaranteed to
satisfy the $\winf$ highest weight conditions of the previous section
that should be checked again.
Here, we cannot provide a general argument, but shall present
some sample calculations that have been done
for $N=3,4,5$, $\ \ k=1,\dots,8$ and low values of the $(n,m)$ indices
with the help of computer algebra.

The check of the incompressibility conditions (\ref{W-inc-nu}) on the shifted 
Laughlin wave function (\ref{red-wf}) gives the following results:
\be
{\cal L}^{(k)}_{nm}\ \Phi_{k-gs}\ =\ 0\ , \qquad {\rm for}\ \ 
0\le n<m=1,2\ ,
\label{W-inc-exp}
\ee
with
\be
{\cal L}^{(k)}_{n1}\ =\ \sum_i\ \l_i^n\ \frac{\de}{\de\l_i} \ ,\qquad
{\cal L}^{(k)}_{n2}\ =\ \sum_i\ \l_i^n\ \left(\frac{\de^2}{\de\l_i^2} \ -
\ \sum_{n,n\neq i}\frac{k^2+k}{(\l_n-\l_i)^2}\right)\ .
\label{W-gen-exp}
\ee
We see that the covariant derivatives are already effective 
for stabilizing the incompressible fluid at second order.
Note the similarities, but also the differences, 
of the covariant derivatives (\ref{W-gen-exp}) and (\ref{W-gen-nu}),
in the present and earlier proposals \cite{flohr}
of $\winf$ generators at fractional filling.
In the eigenvalue representation, both the operators
and the wave functions depend explicitly on $k$: 
this allows us to check the alleged shift $k\to k+1$ in 
$\Phi_{k-gs}=\D(\l)^{k+1}$. 

Next, the operators ${\cal L}^{(k)}_{03}$ and 
${\cal P}^{(k)}_{01}\sim {\cal L}^{(k)}_{01}$
also annihilates the ground state. For higher indices, 
the incompressibility conditions are only satisfied by
specific superpositions of the two kinds of operators (\ref{W-exp}).
For example, we have checked those of:
\ba
\wt{\cal L}^{(k)}_{13}\ =\  {\cal L}^{(k)}_{13}\ +\ \g\ {\cal P}^{(k)}_{02}
\ ,\nl
\wt{\cal L}^{(k)}_{23}\ =\  {\cal L}^{(k)}_{13}\ +\ \s\ {\cal P}^{(k)}_{12}
\ ,
\label{W-inc-mix}
\ea
where $\g(k), \s(k)$ have a non-trivial
dependence on $k$ but are independent of N (as they should):
\be
\begin{array}{c|ccccccc}
k  & 1 & 2 & 3 & 4 & 5 & 6 & 7 \cr
\hline
\g & \frac{1}{4} & 1 & \frac{15}{8} & \frac{14}{5} &
    \frac{15}{4} & \frac{33}{7} &\frac{91}{16} \cr
\s & \frac{1}{2} & 2 & \frac{15}{4} & & & & 
\end{array}\ .
\label{coe-tab}
\ee
The general pattern that emerges from these examples is that the
incompressibility conditions in physical coordinates are satisfied by
specific linear combinations of the
${\cal L}^{(k)}_{nm}$ operators and their finite-N descendants
${\cal P}^{(k)}_{n-s,m-s}$ (that are missing or are too simple in the cases
(\ref{W-inc-exp}).
We guess that the same combinations also obey a closed
$\winf$ algebra.

In conclusion, we have seen that the $\winf$ generators express the
stability (incompressibility) of the Laughlin ground state in 
Chern-Simons matrix model, both in the gauge-invariant and gauge-fixed 
forms. In the latter case, the covariant derivatives are instrumental for
this result and express a form of statistical interaction.


\section{Path integral and integration measure }

\subsection{Real quantization}

In this section we perform the reduction to the physical degrees
of freedom both in the path integral and the state overlaps.
We start by discussing the real case leading to the Calogero
model, basically repeating the analysis in Ref. \cite{gorsky}.

The path integral of the Chern-Simons matrix model is:
\ba
\!\!\!\!\!\!\la f\vert i\ra \!\! & = \!\! & 
\int {\cal D}X_1(t)\ {\cal D}X_2(t)\ 
{\cal D}\psi(t)\ {\cal D}\ov\psi(t)\nl 
&&\qquad\times\exp\int dt\left( -i\ B\ \Tr (X_2 \dot{X}_1) \ +\ 
\psi^\dag \dot{\psi} \ -\ i {\cal H}\right) 
\ \prod_t \ \d(G(t))\ {\rm FP}\ ,
\label{CSMM-real-pi}
\ea
where $G$ is the Gauss-law condition (\ref{gauss-Poly})
and FP is the Faddeed-Popov term for the gauge fixing
(\ref{X-diag}) reducing  $X_1$ to its eigenvalues.
Such gauge fixing can be written in the path integral as follows,
\be
\d\left(\c\right)\ =\ \prod_{i\neq j}\d\left(X^1_{ij}\right)\ =\
\int {\cal D}\L \ \prod_{ij}\d\left(X^1_{ij}-\L_{ij}\right)\ ,
\label{gf-real}
\ee
where $\L=diag(x_1,\dots,x_N)$ is a real diagonal matrix.
The corresponding Faddeev-Popov term is:
\be
1\ =\ FP\ = \int {\cal D}U\ \d\left(\c^U\right)\ 
\det\left(\frac{\de\c^U}{\de\w}\right)\ =\ 
\int {\cal D}U\ {\cal D}\L\ \d\left( U^\dag X_1U-\L\right)\ \D(x)^2\ .
\label{FP-real}
\ee
In this expression, ${\cal D}U$ is the $U(N)$ Haar measure,
$d\w= U^\dag dU$ and the Faddeev-Popov determinant is easily computed to be
the square of the Vandermonde.
Upon inserting (\ref{FP-real}) in the path integral and performing
the gauge transformation, $X_1=U\L U^\dag$, $X_2=U\wt{X}_2 U^\dag$,
$\psi=U\phi$, the Gauss constraint can be rewritten:
\ba
\d\left(G\right)& =& \d\left[U \left( 
B \left[\L,\ \wt{X}_2\right]-i\ k\ I\ +\ i\ \psi\otimes\psi^\dag \right)U^\dag
\right] \nl
& =& \frac{1}{\D(x)^2}\ 
\prod_{i\neq j}\d\left(\wt{X}_{2 ij}+\frac{i}{B}\ 
\frac{\phi_i\ \phi^\dag_j}{x_i-x_j}\right)\ 
\prod_i\d\left(\phi_i\ \phi^\dag_i-k\right)\ .
\label{gf-sol}
\ea
Therefore, we find that the Faddeev-Popov determinant is cancelled
by the Jacobian coming form the solution of the Gauss law.
The conditions (\ref{gf-real},\ref{gf-sol}) do not involve time 
derivatives and 
can be substituted in the path integral (\ref{CSMM-real-pi}) 
at every time step: 
they eliminate the $X_1(t),X_2(t)$ integrations in favor of their 
diagonal elements, $\{x_i(t),y_i(t) \}$.
The $\psi,\psi^\dag$ integrations 
can be performed by substituting $\phi_i=\r_i \exp(i\f_i)$,
using the Gauss law for $\r_i$ and  fixing
the residual $U(1)^N$ symmetry with the linear conditions
$\prod_i\d(\f_i)$ causing no FP determinant.

The kinetic terms in the action can be rewritten using
the constraint (\ref{gf-sol}) as follows:
\ba
-iB\ \Tr\ (X_2 \dot{X}_1)\ +\ \psi^\dag \dot{\psi} &=&
-iB\ \Tr\left(\wt{X}_2\dot{\L} +
[\L,\wt{X}_2]U^\dag\dot{U}\right)\ +\ \phi^\dag\dot{\phi}\ +\ 
\phi^\dag U^\dag\dot{U}\phi \nl
\nl
&=&\sum_n\left(-iB\ y_n\dot{x}_n \ +\ 
\ov{\phi}_n\dot{\phi}_n\right)\ +\ k\ \Tr (U^\dag\dot{U}) \ .
\label{kin-trans}
\ea 
The term proportional to $k$ is a total derivative that 
expresses the variation of
phase of the determinant over the time interval,
$ik\ Arg(\det U)\vert_{t_f}^{t_i}$, and does not contribute to
the path integral for  integer $k$ \cite{nair}.
Therefore, the $U(N)$ integrals factors out and one is left
with the phase-space path integral of the 2N conjugate variables of 
Calogero model, $\{x_i,p_i=-B\ y_i\}$, with Hamiltonian (\ref{S-Calog}):
\be
\la f\vert i\ra \ = \ \int \prod_i\ 
{\cal D}p_i(t)\ {\cal D}x_i(t)\ 
\exp\int dt\left( i\sum_{i=1}^N\ p_i\ \dot{x}_i \ -\ H(p_i,x_i)\right)
\ .
\label{Calog-pi}
\ee

In the real quantization, one is interested in the wave functions
which depend on the coordinate $X_1$ and then, after reduction, on its
eigenvalues. These wave functions can be obtained by replacing
$X=X_1+X_2$ in the the matrix expressions (\ref{HVR-bose}) with
$X_{1ij}$ coordinates and $X_{2ij}$ derivatives w.r.t. them.
In particular, for the ground state, the derivatives vanish 
\cite{karabali} and one recovers the same determinant expression 
(\ref{HVR-bose}) with $X\to X_1$, i.e. $\Psi_{k-gs}(X_1,\psi)$.

The ground-state overlap is defined by (after freezing the vector to
$\phi_i=\sqrt{k}$):
\be
\la\Psi_{k-gs}\vert \Psi_{k-gs}\ra\ =\ \int{\cal D}X_1\ 
{\rm e}^{-B \Tr X_1^2}\ \Vert  \Psi_{k-gs}(X_1)\Vert^2\ .
\label{over-real}
\ee
This expression is actually an Hermitean matrix model, whose
reduction to the eigenvalues is well known \cite{mehta}: nothing depends on
the unitary group that factorizes, leaving the Jacobian $\D(x)^2$ for the
volume of the $U(N)$ gauge group. The wave function becomes
$ \Psi_{k-gs}(U\L U^\dag)=\exp(i\s)\ \D(x)^k \ k^{Nk/2}$, leading to:
\be
\la\Psi_{k-gs}\vert \Psi_{k-gs}\ra\ =\ {\cal N}\ \int{\cal D}\L \ 
\D(x)^2\ {\rm e}^{-B \sum_i x^2_i}\ \D(x)^{2k}\ .
\label{over-gs}
\ee
After reduction, we find that the wave function of the Calogero model 
should be defined with an additional factor of the
Vandermonde, $\Phi_{k-gs}= \D(x)\Psi_{k-gs}$, corresponding
to the shift $k\to k+ 1$ found by several authors 
\cite{poly1,susskind2}.

Another way to understand this shift is through the comparison
of the ground state energies computed in the original matrix theory and
the Calogero model (the case $k=0$ is already significant).
In the Chern-Simons matrix model, the Hamiltonian 
${\cal H}=\w\Tr(X_1^2+X_2^2)$  (setting $B=2$)
is a collection of $N^2$ harmonic
oscillators and the $k=0$ ground state is the Fock vacuum. 
The quantization of these bosonic oscillators 
gives the ground state energy $E_0=\w\ N^2/2$.
On the other hand, the $k=0$ Calogero model contains $N$ oscillators
that would give $E_0=\w\ N/2$ if quantized as boson and 
$E_0=\w\ \sum_{n=0}^{N-1}(n+1/2)=\w\ N^2/2$ if they are fermions.
Therefore, the second choice should be made, leading to the
ground state wave function $\Phi_{0-gs}=\exp(-B\sum x^2_i)\D(x)$ 
as the result of the Slater determinant of the first N harmonic
oscillator states.


\subsection{Holomorphic quantization}

The path integral of the Chern-Simons matrix model 
in holomorphic form (\ref{S-complex}) is:
\ba
\la f\vert i\ra &= & 
\int {\cal D}X(t)\ {\cal D}\ov{X}(t)\ {\cal D}\psi(t)\ {\cal D}\ov\psi(t)\nl
&&\ \ \times\exp\int dt \left(\frac{B}{2}\Tr (X^\dag \dot{X}) \ +\ 
\psi^\dag \dot{\psi} \ -\ i {\cal H} \right)
\ \prod_t \ \d(G(t)) \ {\rm FP}\ .
\label{CSMM-pi}
\ea
The analysis goes in parallel with that of the previous section, with some 
differences regarding the reality conditions.
The classical change of variables that corresponds to the canonical
transformation to the complex eigenvalues\footnote{
Note that the canonical commutators (\ref{X-comm}) are left invariant by
these $GL(N,\C)$ transformations that are more general than the 
$U(N)$ gauge invariance of the complete theory.} 
(\ref{Bogo-trans}) and (\ref{der-trans}), is:
\ba
X &=& V^{-1}\ \L\ V\ ,\qquad\qquad \psi \ =\ V^{-1}\ \phi\ ,\nl
X^\dag &=& V^{-1}\ \wt{\L}\ V\ ,\qquad\qquad \psi^\dag\ =\ \wt{\phi}\ V\ ,
\label{meas-trans}
\ea 
where $\L$ is diagonal and $V$ belongs to the quotient of 
linear complex matrices modulo the
real diagonal ones, $V\in GL(N,\C)/\RR^N$.
In the $k=0$ case, the normal matrices are
diagonalized by a unitary transformation, $V^{-1}=V^\dag$,
thus $\wt{\L}=\L^\dag$ is also diagonal;
for general $k$, the matrix $\wt{\L}$ will be different from 
the conjugate of $\L$ and non-diagonal.
Therefore, we should perform a transformation of the integration
measure in (\ref{CSMM-pi}) that does not respect the complex conjugation
of matrices at the classical level, i.e. a analytic 
continuation of the matrix integral. 
The same remark applies to the integration of $\psi$.
Nonetheless, the final result will be real and well-defined.

In the matrix models arising from $N=2$ topological string theory
\cite{vafa}, one encounters an analogous situation of real (so-called
A-model) and holomorphic (B-model) quantizations.
The holomorphic matrix model has been analysed in depth by
Lazaroiu in  Ref. \cite{lazaroiu}. 
Following his approach, we will consider $X$ and $\ov X$
as independent complex matrices and 
$\int{\cal D}X$ as the holomorphic integral on a curve
$\g\in\C^{N^2}$ that is left invariant by the holomorphic
gauge transformations $V$. The reduction to the eigenvalues is in
this case \cite{lazaroiu}:
\be
\int_\g {\cal D}X\ =\ \int_\g {\cal D}\L\ {\cal D}v \ \D(\l)^2\ ,
\qquad\qquad dv=V^{-1}dV \ .
\label{holo-meas}
\ee
Note that all three terms in the r.h.s. of this equation are holomorphic,
thus respecting the counting of degrees of freedom.

Therefore, the holomorphic version of the Faddeev-Popov term (\ref{FP-real})
is:
\be
1\ =\ 
\int {\cal D}v\ {\cal D}\L\ \d\left(VXV^{-1}-\L\right)\ \D(\l)^2\ ,
\label{FP-holo}
\ee
where the real delta function has been extended to complex holomorphic
arguments.
After the transformation (\ref{meas-trans}) in the path integral,
the Gauss constraint can be solved for $\wt{\L}$ (cf.(\ref{gf-sol})),
leading to:
\ba
 \d\left(G\right)&=&
\prod_{i,j=1}^N\ \d\left(\frac{B}{2}\left(\l_i-\l_j \right)\wt{\L}_{ij}
\ -\ k \ \d_{ij}\ +\ \phi_i\ \wt{\phi}_j \right) \nl
&=&  \D(\l)^{-2}\ 
\prod_{i\neq j}\ \d\left(\wt{\L}_{ij}\ + 
\ \frac{2}{B}\frac{\phi_i\ \wt{\phi}_j}{\l_i-\l_j }\right)\ 
\prod_i\ \d\left(\phi_i\ \wt{\phi}_i \ -\ k\right)\ .
\label{d-meas}
\ea
The Faddeev-Popov determinant and the Jacobian of the
Gauss constraint cancels out in the path integral as in the previous 
case of real quantization.
In the solution $\wt{\L}$, we recognize the off-diagonal
terms of the covariant derivative (\ref{der-trans}):
the complete parametrization is,
\be
\wt{\L}_{ij} \ =\ \wt{\l}\ \d_{ij} \ -\ \frac{2}{B}
\frac{1-\d_{ij}}{\l_i-\l_j }\ \phi_i\ \wt{\phi}_j\ .
\label{lan-til}
\ee
The diagonal elements are unconstrained and become the
canonical conjugate variables of the eigenvalues.
Actually, the substitution of the transformation (\ref{meas-trans})
and of the constraints (\ref{FP-holo},\ref{d-meas}) in the path
integral (\ref{CSMM-pi}), yield the action,
\ba
\frac{B}{2}\ \Tr\ (X^\dag \dot{X})\ +\ \psi^\dag \dot{\psi} &=&
\frac{B}{2}\ \Tr\left(\wt{\L}\dot{\L} -
[\L,\wt{\L}]\dot{V}V^{-1}\right)\ +\ \wt\phi\dot{\phi}\ -\ 
\wt\phi\dot{V}V^{-1}\phi \nl
\nl
&=&\sum_n\left(\frac{B}{2}\ \wt\l_n\dot{\l}_n \ +\ 
\wt{\phi}_n\dot{\phi}_n\right)\ -\ k\ \Tr (V^{-1}\dot{V}) \ ,
\label{kin-holo}
\ea 
and finally (setting $B=2$ hereafter):
\ba
\!\!\!\!\!\!\!\!\!\!\!\!\!\!\la f\vert i\ra & = & \int \prod_n\ 
{\cal D}\wt{\l}_n(t)\ {\cal D}\l_n(t)\ {\cal D}
\wt{\phi}_n(t)\ {\cal D}\phi_n(t)\ 
\prod_{i,t}\d\left( \wt{\phi}_i(t)\ \phi_i(t) - k\right) \nl
&\times&\!\!\!\exp\int dt\left.\left[ 
\sum_n\ \wt{\l}_n\ \dot{\l}_n \ 
+\ \wt{\phi}_n\dot{\phi}_n\ -\  {\cal H}\left(\L,\wt\L \right)
\right] \right\vert_{\wt{\L}_{ij}=
\wt{\l}_i\ \d_{ij}-(1-\d_{ij})\phi_i\wt{\phi}_j/(\l_i-\l_j)}\ . 
\label{qhe-pi}
\ea
The variables $\{\phi_n,\wt\phi_n\}$ are are actually frozen to 
$\sqrt{k}$, $\forall t$, but are left indicated in (\ref{qhe-pi})
to keep track of possible normal orderings in the Hamiltonian.
Once they are eliminated, we recognize that this path integral describes
electrons in the lowest Landau level, with coordinates $\{\l_n,\wt\l_n\}$.
The replacement of $X^\dag$ by $\wt\L$ amounts to a ``twisted'' rule of 
complex conjugation that will better analysed in the next section.

The result (\ref{qhe-pi}) can also be obtained in another way
that uses the Ginibre decomposition of complex matrices 
respecting reality conditions \cite{mehta}.
After diagonalization of $X$, the measure of integration
can be expressed in terms of the eigenvalues and their complex conjugates,
as follows \cite{zabrodin}:
\be
{\cal D}X\ {\cal D}\ov{X}\ {\cal D}\psi\ {\cal D}\ov{\psi}\ =\ 
\left\vert \D(\l) \right\vert^4\ \prod_{i=1}^N\ d\ov{\l}_i\ d\l_i \
 \prod_{i\neq j=1}^N\ dv_{ij}\ d\ov{v}_{ij}\ 
\prod_{i=1}^N\ d\phi_i \ d\ov{\phi}_i \ .
\label{c-meas} 
\ee
The $dv_{ij}$ integral can be further elaborated, 
thanks to the decomposition \cite{mehta}:
\be
V\ =\ D\ Y \ U\ ,
\label{Gin-dec}
\ee
where $D$ is a diagonal, positive real matrix ($D\to I$ for the mentioned
quotient), $Y$ is upper triangular with diagonal elements equal to
one and $U$ is unitary.
The measure $\prod_{i\neq j}\ dv_{ij}\ d\ov{v}_{ij}$
factorized into the unitary measure 
$\prod_{i>j}\ d\w_{ij}\ d\ov{\w}_{ij}$ and the measure
$\prod_{i<j}\ d\a_{ij}\ d\ov{\a}_{ij}$ for $d\a=Y^{-1}dY$.
The earlier variables (\ref{meas-trans})
are expressed in the coordinates (\ref{Gin-dec}) as follows,
\be
\wt{\L}\ =\ H\ \ov\L\ H^{-1}, \qquad\qquad \wt\phi= \phi^\dag H^{-1}\ , 
\qquad\qquad H\ =\ YY^\dag\ ,
\label{Gin-lan}
\ee
and the Gauss constraint reads,
\be
\prod_{i\neq j}\ \d\left( \left(\l_i-\l_j \right)
\left( H \ov\L H^{-1}\right)_{ij}
\ +\ \phi_i\ (\phi^\dag H^{-1})_j \right) \ 
\prod_i\ \d\left(\phi_i\ (\phi^\dag H^{-1})_i \ -\ k\right).
\label{Gin-gau}
\ee
We see that $(N^2-N)$ gauge degrees of freedom have disappeared, while
the $(N^2-N)$ variables of $Y$, i.e. of the similarity transformation $H$, 
should be fixed by the Gauss constraint in terms of 
$\{\l_n,\ov\l_n,\phi_m,\ov\phi_m\}$.

Although in complex conjugate pairs, these coordinates do not have a 
simple dynamics and the reduced path integral cannot 
be interpreted in the quantum Hall effect.
This set of variables was also considered by Karabali and Sakita
\cite{karabali},
with the difference that they allowed $\{H_{nm},\phi_n\}$ to fluctuate because 
no gauge fixing was included in the path integral. 
Although fully legitimate for finite N, this approach 
does not allow to disentangle the
eigenvalues from the remaining variables and makes the physical
interpretation harder.
Therefore, we are led to reintroduce the earlier variables
$\{\wt\l_n,\wt\phi_n\}$ in the path integral via the identities,
\be
1\ =\ \int\prod_n\ d\wt\l_n\ d\wt\phi_n\ 
\d\left(\wt\l_n\ -\ \left(H\ov\L H^{-1} \right)_{nn} \right)\ 
\d\left(\wt\phi_n \ -\ \left(\phi^\dag H^{-1} \right)_{nn} \right)\ , 
\label{add-one}
\ee
and solve for $\{\ov\l_n, \ov\phi_n\}$ and
$\{Y_{nm}\}$ (or $\{H_{nm}\}$ without extra Jacobian, 
$Y^{-1}{\cal D}Y= H^{-1}{\cal D}H$ \cite{mehta}).
The result is the following: the first two deltas in 
(\ref{Gin-gau},\ref{add-one}) combine into 
$\d(\ov\L-H^{-1}\wt\L H)$ times the Jacobian $|\D(\l)|^{-4}$
that cancels the contribution from the integration measure (\ref{c-meas}).
This result is derived in analogy with the other deltas occurring before
and explicitly checked for $N=2$.


\subsection{Complex overlaps}

The integration measure for the Chern-Simons matrix model 
in holomorphic quantization is obtained from the coherent states
of the matrix components (\ref{Fock-meas}), with the inclusion of the
Gauss-law constraint and Faddeev-Popov term ($B=2$):
\be
\left\la\Psi_1 \vert \Psi_2 \right\ra =  
\int {\cal D}X {\cal D}\ov{X}\ {\cal D}\psi {\cal D}\ov{\psi}\  
e^{-\Tr X^\dag X -\psi^\dag \psi}\ 
\d\left(G\right) {\rm FP}\ \
\ov{\Psi_1(X,\psi)}\ \Psi_2(X,\psi)\ .
\label{Fock-OK}
\ee
Using the previous solutions of the constraints (\ref{FP-holo})
and (\ref{d-meas}),  we find the reduced measure:
\ba
\left\la\Psi_1 \vert \Psi_2 \right\ra &=&
\int \prod_n\ d\wt\l_n\ d\l_n\ d\wt\phi_n\ d\phi_n \
e^{-\sum_n \left(\wt\l_n \l_n + \wt\phi_n\phi_n\right) } \nl
&&\ \ \left.\times
\ \ \ov{\Psi_1}(\wt\L,\wt\phi)\ \Psi_2(\L,\phi)
\right\vert_{\wt\phi_i\phi_i=k ,\ \ \wt{\L}_{ij}=
\wt{\l}_i\ \d_{ij}-(1-\d_{ij})\frac{\phi_i\wt{\phi}_j}{\l_i-\l_j}}\ .
\label{over-holo}
\ea
In this expression, the wave function $\Psi_1(X,\psi)$ is first
complex conjugated and then the matrix $X^\dag$ is replaced by
$\wt\L$. The frozen $\{\wt\phi_n,\phi_n\}$ are again
maintained for allowing normal orderings.
For example, the $N=2$ ground state overlap reads (up to constants):
\be
\left\la\Psi_{k-gs} \vert \Psi_{k-gs} \right\ra_{N=2} =
\int  d\wt\l_1\ d\l_1 d\wt\l_2\ d\l_2\ 
e^{- \wt\l_1 \l_1 -\wt\l_2 \l_2  } \ 
\left( \wt\l_1 - \wt\l_2 \ + \ \frac{2k}{\l_1-\l_2} \right)^k
\ \left(\l_1-\l_2\right)^k\ .
\label{gs-2}
\ee
Owning to the twisted rule of conjugation, 
this overlap does not have the standard expression for
the Laughlin wave function in the quantum Hall effect:
\be
\left\la\Psi_{k-gs} \vert \Psi_{k-gs} \right\ra_{N=2} =
\int  d\ov{z}_1\ dz_1 d\ov{z}_2\ dz_2\ 
e^{- \ov{z}_1 z_1 -\ov{z}_2 z_2  } \ 
\left( \ov{z}_1 - \ov{z}_2 \right)^k\ \left(z_1-z_2\right)^k\ .
\label{gs-qhe}
\ee
Here, we remark that the physically relevant quantities are the
values of the overlap integrals, normalized to that of the ground state:
thus, the discrepancy between (\ref{gs-2}) and (\ref{gs-qhe})
does not immediately imply that the matrix theory does not
describe the Laughlin states.

The rule of conjugation in the overlap integral (\ref{over-holo}) 
is consistent with the result of the change of variables of
section 3: actually, we see that the covariant derivative (\ref{der-trans})
acting on the holomorphic wave function $\Psi_2$, 
can be adjoined as follows:
\be
\la \Psi_1\vert Tr (X^{\dag n})\ \Psi_2\ra\ =\ 
\la Tr (X^n)\ \Psi_1\vert\Psi_2\ra\ .
\label{adj}
\ee
Therefore, the $\winf$ operators satisfy the correct Hermiticity rule,
${\cal L}_{nm}^\dag = {\cal L}_{mn}$, both in
the matrix (\ref{W-mat}) and reduced (\ref{W-exp}) coordinates.
In conclusion, the twisted conjugation rule
guarantees the Hermiticity of the covariant derivative that
shows up in the reduced coordinates.

This result is rather important for the physical interpretation
of the Chern-Simons matrix model.
The main physical aspect of this theory is the 
realization of the Laughlin wave function in the ground state
thanks to the conditions set by the Gauss law,
following from the classical non-commutativity of fields (section 3.1).
Actually, $U(N)$ group theory arguments \cite{heller}
prove that all wave functions should contain the Laughlin
ground state as a factor (cf. Eq.(\ref{HVR-bose})), 
thus the ground state is stable (incompressible).
Unwanted states representing smaller droplets of fluid 
are possible in the matrix Fock space, but they 
do not respect the Gauss law and are not physical states.
In conclusion, stability come from gauge invariance and 
the Gauss law condition.

In the reduced eigenvalue basis, the gauge symmetry has been
projected out, but the stability of the ground state is still
proven the $\winf$ incompressibility conditions, that
are satisfied thanks to the statistical
interaction, i.e. to the covariant derivatives.
As in the Chern-Simons-matter field theory approach by 
Fradkin and Lopez \cite{fradkin}, the stability
of the Laughlin ground state is realized by adding a
further gauge interaction, rather than a two-body repulsion
between the electrons \cite{haldane}.
In the reduced theory (\ref{qhe-pi},\ref{over-holo}),
ordinary derivatives $\de/\de\l_n$ could create 
states of lower energy and higher density, i.e. generate instabilities;
however, these operators are forbidden because they are not Hermitean
under the twisted conjugation rule 
(\ref{over-holo}), and thus create non-unitary states;
the $\winf$ generators ${\cal L}_{nm}$ 
are the only available Hermitean operators for creating excitations.
In conclusion, the Hermitean conjugation of (\ref{over-holo})
is consistent with the presence of the statistical interaction
that stabilizes the Laughlin ground state.

It remains to be proven that the unconventional overlaps (\ref{over-holo})
take the same value (for physical states only) of the ordinary
quantum Hall expressions.
Actually, as discussed in section 4.1, Eq.(\ref{W-over}),
a coordinate-free characterization of the overlaps 
in the quantum Hall effect is obtained by expressing them 
as commutators of the $\winf$ algebra.
Thus, the proof of the correspondence between the matrix and Hall overlaps
is equivalent to the derivation of
the unitary representations of the $\winf$ algebra
in the Chern-Simons matrix model.

Another aspect of measure of integration (\ref{over-holo})
is the lack of the shift $k\to k+1$, that was found in the real 
quantization and was crucial for the analysis of the $\winf$ symmetry 
in section 4.3.
In particular, for $k=0$, we do not recover 
the known result\footnote{
Note that $\wt\l_n\to\bar\l_n$ for $k\to 0$.} 
$|\D(\l)|^2$ for the measure of 
the ensemble of normal complex matrices \cite{mehta}, that should
correspond to the $\nu=1$ quantum Hall effect \cite{zabrodin}.
The reason of this discrepancy is coming from the different
definitions of the measure of integration: in our case, we solved
the delta function of the Gauss constraint coming from the
integration of the $A_0$ field in the Lagrangian; in Ref. \cite{zabrodin}, 
they computed the induced metric on the manifold of the classical constraint 
$[X,X^\dag]=0$ in the space $\C^{N^2}$ of complex matrices.
The two analyses are not in contradiction: indeed, the Gauss constraint 
is degenerate for $k=0$, where it admits both types of solution, 
depending on how the quantity $\d(0)$ is regularized. 
On the contrary, for $k>0$, the constraint is made non-degenerate 
by the presence of the extra field $\psi$ and our results are
not ambiguous and hold in the limit $k\to 0^+$.

The study of the ground state energy provides another way to analyse
this issue, as discussed at the end of section 5.1. 
In the present case of quadratic Hamiltonian, the ground state energy 
is also related to the filling fraction, i.e. to the size of the 
droplet of fluid \cite{susskind2}:
for a circular droplet of uniform fluid, one finds 
$\la R^2\ra = (N/2\pi)\ Area$, and therefore:
\be
\frac{1}{\nu}\ = \ \frac{B\ Area}{2\pi\ N}\ = \
\frac{B}{N^2}\ \la\Tr\left( X_1^2+X_2^2\right)\ra\ =
\ \frac{2\w}{N^2}\ E_0\ .
\label{eff-fill}
\ee
In the real case, we got $\la\Tr\left( X_1^2+X_2^2\right)\ra
=\la\Tr\ X\ X^\dag +N^2/2\ra$, namely
a shift in the ground state energy leading to $1/\nu=k+1$.
In the holomorphic matrix quantization of section 3, the
normal ordering is 
$\la \Tr\ X^2_a\ra=\la \sum_{ij} X_{ij}\ \de/\de X_{ij} \ra $,
yielding no shift (and vanishing $E_0$ for $k=0$).
Thus, it seems that our results (\ref{over-holo}) is consistent
with the operator ordering of matrices in holomorphic quantization.
On the other hand, the ordering adopted for
the $\winf$ operators in the reduced coordinates (\ref{W-exp})
seems to be different, because the incompressibility conditions
are satisfied by the shifted wave function $\Phi_{k-gs}$ 
(cf. Eq.(\ref{W-inc-exp})).
The shift in the wave functions is also important 
for changing the statistic of the reduced particle degrees
of freedom from bosonic to fermionic. 

Although we do not presently understand these facts completely,
they seem to indicate that the derivation of the reduced
holomorphic path integral and overlap should be modified
by allowing for ground state fluctuations.
One possibility is to include an additional measure by hand
in the starting expression of the complex matrix overlap (\ref{Fock-OK}).
This should be a gauge invariant, self-adjoint (real positive) quantity, 
carrying no charge for the $\psi$ field, i.e. made of $X,X^\dag$ only,
and should reduce to $|\D(\l)|^2$ for $k=0$.
The following measure satisfies these rather stringent requirements and
is inspired by the work of Ref. \cite{callaway}:
\be
{\cal M}\ = \ \det\left(\Tr\left(X^{\dag i-1}\ X^{j-1}\right)\right)\ 
=\ \det\left(\left(\wt{\L}^{i-1}\right)_{jj} \right)\ \D(\l)\  .
\label{guess-meas}
\ee
This quantity inserted in the overlap (\ref{Fock-OK}) realizes the shift
$k\to k+1$ in the eigenvalue wave function $\Phi$ of section 4.3.
Note also that a similar determinant expression involving the $\psi$ field,
$\det\left(\psi^\dag\ X^{\dag i-1}\ X^{j-1}\ \psi)\right)$,
is actually the modulus square of the ground state wave function.
The measure (\ref{guess-meas}) should also be included in 
the holomorphic path integral (\ref{CSMM-pi}) for consistency, 
where it defines an instantaneous repulsive interaction that enforces
the fermionic character of the electrons.


\section{Conclusions}

In this paper, we described the holomorphic quantization of the
Chern-Simons matrix model; we performed a change of matrix
variables that allowed to solve the Gauss-law constraint and
to map the $N^2$-dimensional quantum mechanics of D0 branes 
with Chern-Simons dynamics into the problem 
of $N$ particles in the lowest Landau level. 
We found that the matrix theory describes the Laughlin 
ground state wave function and the corresponding excitations,
together with a statistical interaction that is crucial to
stabilize the ground state and verify the $\winf$ incompressibility
conditions.
Some gaps are still present in our analysis: we did not obtain
the full representation of the $\winf$ algebra in the theory
and could not completely determine the measure in the overlap integrals.
Nonetheless, we showed that the eventual proof of the $\winf$ 
symmetry in the matrix theory would definitely establish the correspondence 
with the Laughlin Hall states, independently of coordinate choices.

The non-commutative Chern-Simons theory has shown rather interesting features:
besides incorporating the discrete nature of the electrons
\cite{susskind}, it enhances the quantum repulsion of 
the electrons in the first Landau level,
\be
[[\ov{\l}_n,\l_m]]\ =\ \frac{2\hbar}{B}\ \d_{nm} \ , 
\ee
by means of the non-commutativity of classical matrices,
\be
[X,X^\dag ]\ = \ 2\th \ ,
\label{theta-comm}
\ee
thus reducing the filling fraction from $\nu=1$ to $\nu=1/(1+B\th)$.
The (integer) statistical interaction among the electrons found 
in holomorphic quantization can actually be deduced from the 
non-commutativity. 
Equation (\ref{theta-comm}) can be interpreted as a gauge field strength
by representing non-commutative fields, i.e. matrices, as covariant 
derivatives. Setting $X=Z+R$, where $Z$ is a normal
matrix (unitarely equivalent to $\L$), we can view $Z$ as 
the ordinary derivative ($[Z,Z^\dag]=0$) and $R$ as the gauge
potential. The solution of (\ref{theta-comm}) found in section 3 is thus:
\be
X\ =\ Z\ ,\quad (R = 0)\ ,\qquad X^\dag \ = Z^\dag\ +\ R^\dag\ ,\qquad
[Z,R^\dag]\ =\ 2\th \ .
\label{stat-int}
\ee
Note that such decomposition of the covariant derivative
is preserved under $GL(N,\C)$ transformations $X\to V^{-1} X V$ provided 
that the gauge potential transforms as
$R\to V^{-1} R V +V^{-1} [Z,\ V]$.

A known property of the Laughlin Hall states is the presence of 
non-trivial quantum-mechanical long-range correlations that cause the 
fractional statistics of excitations \cite{wilczek} and the
topological order of the ground state on compact surfaces \cite{wen}.
An interesting open question is how to recover these features in
the matrix theory.
The $1/(k+1)$ fractional statistics of two quasi-hole excitations, with
matrix wave function,
\be
\Psi_{\rm 2q-hole}(z_1,\ z_2\ ; X,\psi) \ = \ \det\left(z_1 - X\right)
\ \det\left(z_2 - X\right)\ \Psi_{k-gs}(X,\psi)\ ,
\label{q-hole}
\ee
should come out as a (re)normalization effect in the limit $N\to\infty$.
Note that for N finite this state belongs to the same (unique) $\winf$ 
representation that includes the ground state (cf. Eq.(\ref{HVR-bose})).

The Chern-Simons matrix model could provide a concrete
setting for studying the ``exactness'' and ``universality'' of
the Laughlin wave function, upon using
the sophisticated tools of matrix models \cite{zabrodin} \cite{wiegmann},
as well as the new features of non-commutative field theories
\cite{NCFT}\cite{ncgeom}.
Among the possible developments, we mention:
\begin{itemize}
\item
The description of the Jain states at filling fraction
$\nu=\left(k+1/m\right)^{-1}$ by a suitable extension of 
the boundary terms (cf. Ref.\cite{poly3}).
\item
The study of Morita's equivalences in the non-commutative theories
\cite{NCFT}\cite{ncgeom}\cite{barbon},
that change the non-commutativity parameter $\th$ by
a $SL(2,\Z)$ transformation, and 
could actually relate the matrix theories of different plateaus.
\end{itemize}

\bigskip

{\large \bf Acknowledgments}

We would like to thank A. Polychronakos, D. Seminara, P. Wiegmann 
for interesting discussions. We also thank G. Zemba for fruitful
collaboration and D. Karabali for pointing out some problems in the first
version of the paper.
A. Cappelli thanks LPTHE, Jussieu, the Schr\"odinger Institute, Vienna,
and the PH-TH Division, Cern, for hospitality.
This work was partially funded by the EC Network contract 
HPRN-CT-2002-00325, 
``Integrable models and applications: from strings to condensed matter''.


\end{document}